\documentclass[sigconf]{acmart}

\usepackage{booktabs} 
\usepackage{courier}
\usepackage{epsfig}
\usepackage{endnotes}
\usepackage{verbatim}
\usepackage{subcaption}
\usepackage{setspace}
\usepackage{amsmath}
\usepackage{multirow}
\usepackage{xspace}
\usepackage{enumitem}
\usepackage{graphics}
\usepackage{hhline}
\usepackage{ctable}
\usepackage{multirow}

\renewcommand\footnotetextcopyrightpermission[1]{} 
\setcopyright{none}

\acmDOI{10.475/123_4}

\acmISBN{123-4567-24-567/08/06}

\acmConference[DAC'18]{Design Automation Conference}{June 2018}{San Francisco, California USA} 
\acmYear{2018}
\copyrightyear{2018}

\acmArticle{6}
\acmPrice{15.00}

\acmSubmissionID{123-A12-B3}

\newcommand{\tR}{$t_{R}$\xspace}
\newcommand{\tPROG}{$t_{PROG}$\xspace}
\newcommand{\tDMA}{$t_{DMA}$\xspace}
\newcommand{\tDMAIn}{$t_{DMA^{in}}$\xspace}
\newcommand{\tDMAOut}{$t_{DMA^{out}}$\xspace}
\newcommand{\tCOPY}{$t_{COPY}$\xspace}
\newcommand{\rcFTL}{{\small{\sf rcFTL}}\xspace}

\newcommand{\RcFTL}{{\small{\sf RcFTL}}\xspace}
\newcommand{\rcopyback}{{\tt rcopyback}\xspace}
\newcommand{\Rcopyback}{{\tt Rcopyback}\xspace}
\newcommand{\degree}{\ensuremath{^\circ}}

\begin{document}

\title{Revitalizing Copybacks in Modern SSDs:  Why and How}

\author{Duwon~Hong,
        Myungsuk~Kim,
        Jisung~Park,
        ${^\dagger}$Myoungsoo~Jung,
        and~Jihong~Kim, \\
    	\vspace{1pt}
	    \textit{Seoul National University and ${^\dagger}$Yonsei University}
	}
	    

\begin{abstract}
For modern flash-based SSDs, the performance overhead of internal data migrations 
is dominated by the data transfer time, not by the flash program time as in old SSDs.   
In order to mitigate the performance impact of data migrations, 
we propose \rcopyback, a restricted version of copyback.
\Rcopyback works like the original copyback except that only $n$ consecutive copybacks are allowed.  
By limiting the number of successive copybacks, it guarantees that no data reliability
problem occurs  when data is internally migrated using \rcopyback.
In order to take a full advantage of \rcopyback,
we developed a \rcopyback-aware FTL,  {\scriptsize{\sf rcFTL}},
which intelligently decides whether \rcopyback should be used or not by exploiting varying host workloads.
Our evaluation results show that {\scriptsize{\sf rcFTL}} can improve the overall I/O throughput 
by 54\% on average over an existing FTL which does not use copybacks.  
\end{abstract}

%
%

\keywords{Copyback, NAND flash memory, FTL, Storage system}

\maketitle

\section{Introduction}\label{sec:introduction}
Flash-based SSDs move a large amount of data internally for supporting
various SSD management tasks such as garbage collection
(GC), wear leveling and reliability enhancement. For example,
because of the erase-before-write constraint in the NAND flash
memory, GC is required to reclaim invalid pages for
future writes. During GC, valid pages of a GC victim
block should be migrated to a new target block with free pages.
Since these internal data copy operations directly interfere with I/O
requests from user applications, how to efficiently handle internal
data migrations is a key challenge for designing a high-performance
SSD.

Although there have been extensive investigations (e.g.,
~\cite{gupta2009dftl,kang2006superblock, kim2002space, lee2008last, lee2006fast})
to mitigate the impact of internal data migrations on the SSD performance, 
most existing techniques do not adequately handle a new performance bottleneck
of copy operations in modern SSDs.  Unlike old SSDs
where the copy cost was dominated by the flash program time \tPROG, in
recent high-end SSDs, the data transfer time \tDMA between flash cells and
off-chip DRAM takes a large portion of the copy cost.  This shift in the
performance bottleneck is due to two recent flash/SSD technology changes: 
1) innovations in the flash
cell design (which reduced \tPROG)~\cite{lee20167} and 2) a high degree of the internal
parallelism in high-end SSDs (which results in frequent access collisions
on a shared medium (e.g., a channel bus or a serial DRAM bus) between flash cells and off-chip DRAM.)

In order to minimize \tDMA, copyback operations~\cite{IDM} are the most
effective solution because the copyback operation can move pages within an SSD
without off-chip data transfers, thus eliminating \tDMA completely.  However,
copyback operations are rarely used in modern SSDs because they cause a fatal
reliability problem. When pages are migrated using copyback operations, they
bypass an off-chip error-correction code (ECC) module and bit errors occurred
during copyback operations are accumulated.  If the number of the accumulated
bit errors exceeds the correction capacity of the ECC module, the stored data
in the copybacked page becomes {\it unreadable}.  Furthermore, since \tPROG was
responsible for a large portion of the data migration time in old SSDs, the
performance improvement from copyback operations was marginal.  In this paper,
however, we argue that it is time to revitalize old copyback operations for
modern SSDs.

We revisit copyback operations in the context of modern high-density flash
memory and propose a restricted version of copyback, called {\tt rcopyback},
which works like the original copyback except that data migrated by $n$ successive 
{\tt rcopyback} operations must be error-corrected by an off-chip ECC module. 
The proposed {\tt rcopyback} technique is based on a simple observation on the error propagation
characteristics of successive copyback operations.  From our characterization
study with recent 1{$\mathsf{x}$} nm-node NAND chips, we observed that if we properly limit
the number of {\it consecutive} copyback operations, accumulated bit errors can
be within the error correction capability of a common ECC scheme.  Furthermore,
we observed that the overhead of internal data migrations is significantly
reduced even when only a small number of copyback operations can be
successively used. For example, when only two consecutive copyback operations
are possible, \tDMA is effectively reduced by 1/3.

Based on the {\tt rcopyback} model from a detailed characterization study using
1{$\mathsf{x}$} nm-node NAND MLC chips, we designed a new FTL, called \rcFTL, which takes
advantage of {\tt rcopyback} for data migrations.  In addition to basic
extensions for supporting {\tt rcopyback}, \rcFTL implements an intelligent
data-migration mode selector for maximizing the effect of {\tt rcopyback} on
the SSD performance.  The mode selector decides whether {\tt rcopyback} or an
off-chip copy operation is used for a given data migration.  For light-load
intervals, \rcFTL uses the off-chip copy mode, which increases the
number of future \rcopyback-eligible blocks.  On the other hand, for
heavy-load intervals, \rcFTL maximally utilizes {\tt rcopyback} for higher I/O
performance. 

We have evaluated \rcFTL using various benchmarks on our SSD emulation
environment~\cite{jun2015bluedbm}.   Our experimental results show that \rcFTL
can improve the overall IO throughput by 54\% on average over a conventional FTL
without copyback support. We  also  show  that the proposed migration mode selector
is effective in maximizing the efficient use of \rcopyback under varying workload requirements.

\section{Motivations}
\begin{figure}[t!]
    \centering
    \includegraphics[width=\columnwidth]{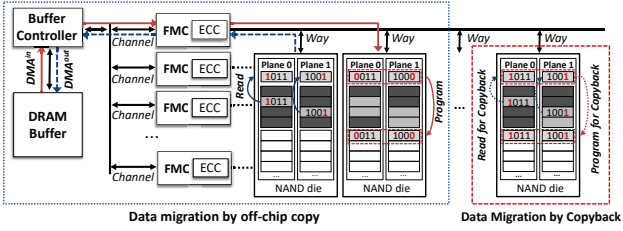}
    \caption{A data path comparison between an off-chip data migration and an internal copyback.}
    \label{fig:data_path_comparison}
\end{figure}

A typical data migration in SSDs is performed by an off-chip data copy 
as shown in the left dotted box of Fig. \ref{fig:data_path_comparison}. 
An SSD firmware reads data from a source page and transfers the data to
a DRAM buffer through a channel bus. 
Before the data are sent to the DRAM buffer, errors are corrected by 
the ECC module of the flash memory controller (FMC).
In the program phase, in order to move the data back to the target page, 
the SSD firmware takes a reverse data path from the DRAM buffer to the target page.
When no contention occurs along the off-chip copy data path,  
the data copy latency \tCOPY can be expressed as follows: 
\tCOPY = \tR + \tDMAOut + \tDMAIn + \tPROG 
where \tR, \tDMAOut and \tDMAIn are a data transfer
time from NAND cells to a per-plane register and a DMA out/in time between the
register and DRAM buffer, respectively.

However, in a modern SSD which consists of multiple channels and multiple NAND dies per channel, 
a large number of data migrations may occur at the same time.  
A high degree of the parallelism in data migrations may significantly increase  
\tDMAIn and \tDMAOut because of contentions on the channel level as well as 
the serial bus to/from the DRAM buffer.
For example, when eight data migrations are concurrently requested by the SSD firmware, 
if all eight migrations had both the source page and destination page on the same channel, 
\tDMAIn and \tDMAOut may increase by eight times because all data transfers should be serialized.

On the other hand, when a copyback command is supported by a NAND flash chip, 
a data migration can be performed without requiring neither \tDMAOut nor \tDMAIn 
as shown in the right dotted box of Fig. 1. 
The SSD firmware can read data from the source page to the per-plane local register 
and directly write back to the destination page from the per-plane local register.   
Since the copyback command transfers data within a given plane, 
even when multiple data migrations occur at the same time, 
if they can be supported by the copyback command, 
all data migrations can complete by (\tR + \tPROG).
Thus, if the copyback command can be supported, it can significantly reduce 
the overhead from SSD-internal data migrations.   
Unfortunately, however, the copyback command is rarely used in modern SSDs 
because it accumulates all the bit errors occurred to a page during its migrations. 
Since, in older NAND flash memory, \tPROG, which is much larger than \tDMA, dominated \tCOPY, 
little effort was made to overcome the error propagation problem of the copyback command. 
\begin{figure}[t]
    \centering
    \includegraphics[width=0.8\columnwidth]{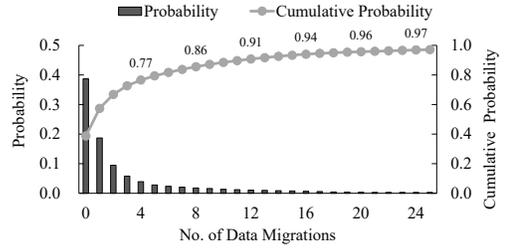}
	\caption{A probability distribution of internal data migrations.}
    \label{fig:dataMigration}    
\end{figure}

In order to develop an effective solution to revitalize the copyback for modern SSDs, 
our proposed technique is motivated by an observation on internal data migration 
characteristics of storage workloads: 
{\it most pages migrate internally just a few times.}  
For example, Fig. 2 shows a probability distribution of 
internal data migrations in RocksDB under the append-random workload of db\_bench.
77\% of pages migrate less than five times.   
Therefore, if we could support 4 consecutive copybacks
without causing any flash reliability problem, 
about 86\% of off-chip data migrations can be avoided in this workload. 

\section{Rcopyback: Copyback with a Threshold}
\label{section:error_propagation}

\subsection{Copyback  Error-Propagation Characteristics}

In order to manage the flash reliability problem caused by successive copyback operations, 
it is important to understand the error propagation characteristics when 
the same page experiences consecutive copyback operations without the ECC module's involvement.
Using 1{$\mathsf{x}$} nm-node MLC NAND chips, we conducted experiments using a total of 81,920 pages out of 20 NAND chips.   
First, we confirmed that, in our tested MLC NAND blocks, which consist of 64 word lines 
(WLs, each of which can store two pages, the MSB page and LSB page), MSB pages of WL 62\footnote{In fact, WL 63 is the most unreliable WL. 
Because of the reliability problem, WL 63 is configured to work as SLC cells.}  were the most unreliable 
because 
the outer WLs, the more vulnerable to noise (e.g., hot-carrier effects and gate-induced drain leakage (GIDL))  and the more disturbed from Vpass.  
Since we are interested in finding a safe bound on the number of consecutive copybacks over all the possible data migrations,
we focus on understanding the error propagation characteristics when both the source and destination pages are
in WL 62, which is the worst combination from the bit-error rate (BER) perspective.

As with other NAND flash reliability evaluations, 
we used the NAND retention BER as our measurement metric.   
(The NAND retention BER is based on the number $N(x, t)$ of bit errors 
after $t$-month retention time at 30\degree C for $x$ pre-cycled NAND cells~\cite{kim2017improving}.) 
For a given upper limit on the number of consecutive copyback operations, we measured $N(x,t)$ values while changing both $x$'s (i.e., P/E cycles) and $t$'s (i.e., retention times).  
\begin{figure}[!t]
	\centering
	\begin{subfigure}[t]{.245\textwidth}
		\centering
		\includegraphics[width=1.0\linewidth]{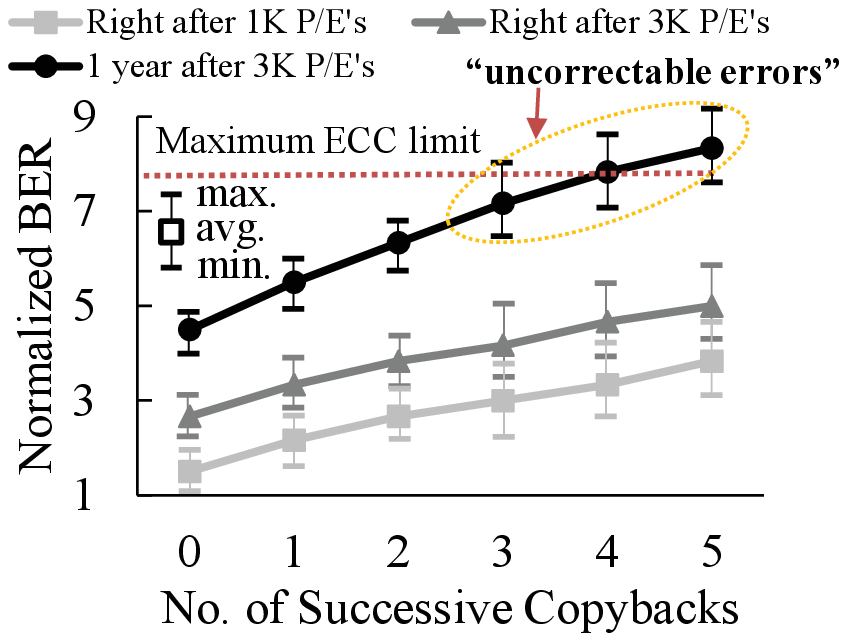}
		\caption{BER variations over successive copybacks under varying P/E cycles.}
	\end{subfigure}%
	\hfil
	\begin{subfigure}[t]{.21\textwidth}
		\centering
		\includegraphics[width=1.0\linewidth]{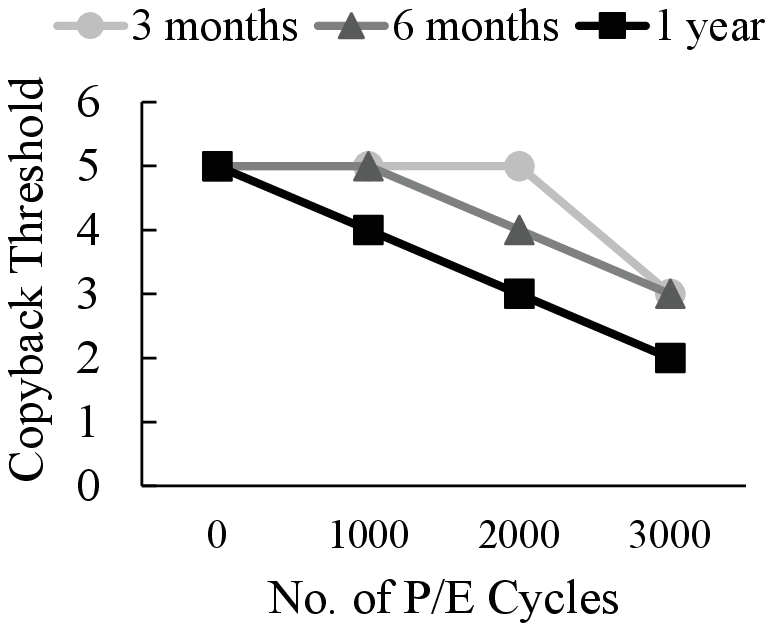}
		\caption{Changes in copyback threshold values under varying P/E cycles.}
	\end{subfigure}%
	\caption{The effect of successive copybacks on the reliability.}
	\label{fig:nand_trend}
\end{figure}
Fig. 3(a) shows how retention BER's change as the number of successive copybacks increases.  
Retention BER values were normalized over $N(0, 0)$.  
For example, $N(3K, 1$ $year)$ values increase almost linearly over the number of consecutive copybacks.   
When a block is erased by 3,000 times, if the 1-year data retention time is required, 
copyback operations cannot be successively used for more than twice.

\subsection{Rcopyback Operation Model} 
From our characterization study on the copyback error propagation, 
we constructed a table of copyback threshold $CT(x, t)$ values for $x$ P/E-cycled NAND blocks with $t$-month retention requirement.   
The $CT(x, t)$ value indicates the maximum number of consecutive copyback operations 
that does not cause any reliability problem for $x$ P/E-cycled blocks when $t$-month data retention is required.  
Fig. 3(b) shows  how copyback threshold values change under varying P/E cycles 
and different retention time requirements for our evaluated MLC NAND chips. 
As the retention time requirement increases, the copyback threshold decreases for the same P/E cycle. 
For the same retention time requirement, as expected, the number of P/E cycles strongly affects the copyback threshold.
For the 1-year data retention requirement at 30\degree C (which is the JEDEC client class retention requirement), 
the copyback threshold value decreases from 5 to 2 as the P/E cycle increases from 0 to 3,000.
That is, after 3K P/E cycles, the copyback command can be consecutively used only twice.  
If the third data migration is required on the same page, 
the page must be migrated using an off-chip data copy, 
thus the accumulated bit errors can be corrected by the ECC engine of the FMC.
\begin{table}[b]
	\centering
	\caption{\Rcopyback operation model.}
    \footnotesize
	\begin{tabular}{ c | c | c | c }
		\specialrule{.1em}{.05em}{.05em}
		  P/E cycle  & {\sf 1-1000} & {\sf 1001-2000} & {\sf 2001-3000} \\
		\hhline{====}
		 Copyback Threshold & 4 & 3 & 2 \\
		\specialrule{.1em}{.05em}{.05em}
	\end{tabular}
	\label{tab:threshold}%
\end{table}%
Table \ref{tab:threshold} summarizes our proposed \rcopyback operation model with the 1-year data retention requirement 
based on the copyback threshold values described in Fig. 3(b).

\section{Design and Implementation of rcFTL} 

\begin{figure}[t]
	\centering
    \includegraphics[width=0.9\columnwidth]{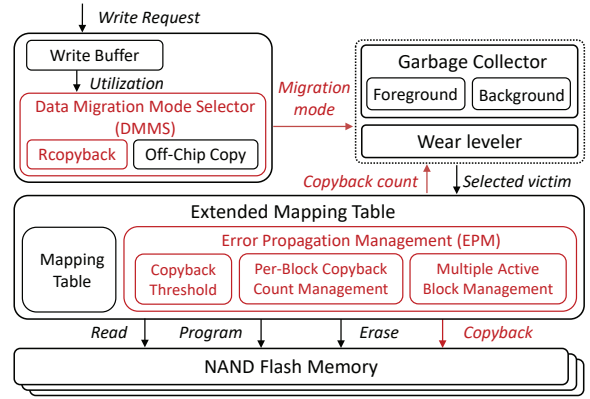}
    \caption{An organizational overview of \rcFTL.}
    \label{fig:ftl_overview}    
\end{figure}

Based on the proposed {\tt rcopyback} model presented in Section 3, 
we implemented \rcFTL which efficiently uses {\tt rcopyback} operation for
data migrations. Fig. \ref{fig:ftl_overview} shows an overall organization of \rcFTL.
\RcFTL, which is based on an existing page-level mapping FTL, consists of two additional modules, 
the error propagation management (EPM) module and the data migration mode selector (DMMS) module.  
The EPM module efficiently manages various data structures for supporting \rcopyback  
while the DMMS module selects the most appropriate data copy mode for a given data migration request.

\subsection{Error Propagation Management}

The main function of the EPM module is to monitor 
the cumulative number of successive copyback operations 
for each page so that no page can be {\tt rcopyback}ed more than the copyback threshold.
A simple approach to keep track of the cumulative count is to maintain a per-page counter 
which is incremented whenever \rcopyback is used for the page.  
However, for recent high-capacity NAND flash memory, the space overhead of per-page counting is quite high.  
For example, 1.4-GB memory is needed for supporting a 3-bit per-page counter for a 16-TB SSD.  
In addition, in highly-optimized commercial SSDs, updating the per-page counter (in slower DRAM memory) 
can incur a significant CPU cycle overhead as well because 
memory accesses are optimized to occur in SRAM memory for higher performance.
In order to avoid the overhead of per-page counting, 
the EPM module employs a {\it per-block} counting approach.  
That is, the cumulative number of \rcopyback operations is manged at the block level, not at the page level.  
Since the number of counters for the per-blocking counting is at least two orders of magnitude smaller than
that for the per-page counting, the per-block counting technique significantly reduces the memory footprint for maintaining counters and minimizes the computing overhead of bookkeeping operations to a negligible level.   

\begin{figure}[t]
	\centering
    \includegraphics[width=0.8\columnwidth]{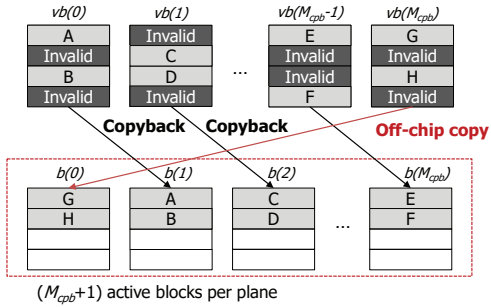}
    \caption{Data migrations in the per-block counting scheme.}
    \label{fig:mutlipleActiveBlock}    
\end{figure}

Since all the pages in a block are assumed to have been migrated by the same number of 
\rcopyback operations in the per-block counting scheme,  
when a source page $p$ in a block $b(c)$ with the counter value $c$ is migrated by \rcopyback, 
the page $p$ should be moved to a page in a block $b(c')$ where $c' = c + 1$. 
In order to efficiently support this additional constraint, 
the EPM module manages {\it multiple active blocks} at the same time.  
If the maximum copyback threshold value is given by $M_{cpb}$, the EPM module maintains 
($M_{cpb} + 1)$ active blocks, ${b_{0},..., b_{M_{cpb}}}$, where $b_{i}$ indicates a block with its counter value $i$.  
Fig.~\ref{fig:mutlipleActiveBlock} shows an example of 
how data migrations are performed using \rcopyback operations in the per-block counting scheme.   
For example, when the block $vb(1)$ is selected as a GC victim block, its valid pages, 
C and D, are moved to the active block $b(2)$ when they are migrated using \rcopyback operations.  
When the block $vb(M_{cpb})$ is selected as a GC victim block, its valid pages are moved 
using off-chip copies to the active block $b(0)$.

\subsection{Data Migration Mode Selection}

Since the copyback threshold is rather small, using \rcopyback in a greedy fashion may not be 
the most effective use of it from an overall I/O throughput perspective.   
For example, when no high I/O throughput is required, it does not make sense to use \rcopyback for data migrations.  
Doing so may prevent more effective future use of \rcopyback when high I/O throughput is needed.   
Furthermore, when the high I/O bandwidth is not necessary, using off-chip data migrations enables 
more future data migrations to be supported by \rcopyback. 

In order to take full advantages of \rcopyback, the DMMS module 
intelligently chooses when to use \rcopyback operation 
over a normal off-chip copy depending on the write buffer utilization ratio $u$.
  When $u$ is low, which indicates that the current host I/O workload is not intensive, 
the DMMS module selects the off-chip copy mode 
so that more future data migrations can be supported by \rcopyback.
On the other hand, when $u$ is high, the DMMS module chooses the \rcopyback mode for higher performance.  
In our current implementation, the utilization threshold ratio for the mode selection was set to 50\%.  
(That is, if $u$ is higher than 50\%, the \rcopyback mode is used for data migrations.)
Since \rcFTL employs the per-block counting scheme and 
most data migration decisions are made in a block granularity, 
the DMMS module makes its mode selection decisions in a per-block level as well.   
When a data migration decision is made (e.g., by a foreground GC task), 
the DMMS  module selects a proper mode based on the current $u$ value.  
In order to filter out abrupt noise-like changes in $u$, 
the DMMS module makes its mode selection based on a $t$-second moving average of $u$.  
In the current implementation, $t$ is set to an average block write time.  

In \rcFTL, both the garbage collector and wear leveler operate in an \rcopyback-aware fashion.  
For urgent management tasks (such as a foreground GC task), the \rcopyback mode is actively used regardless of the current $u$ ratio value.  On the other hand, when background management tasks (such as a background GC task) are invoked, the DMMS module decides proper modes as explained above.

\section{Experimental Results}

\subsection{Experimental Setup} 
In order to evaluate the effectiveness of the proposed \rcFTL technique, 
we implemented \rcFTL as a host-level FTL on a custom flash storage system~\cite{jun2015bluedbm}. 
For our evaluation, we configured our flash storage system to support a 64-GB storage capacity only for efficient experimental evaluations.
Our emulated storage system was configured to have eight channels 
with eight NAND flash chips per channel.  
Each NAND flash chip has 1024 blocks which are composed of 64 16-KB pages.  
The average \tPROG was set to 640 us and the size of the write buffer was set to 10 MB.
We evaluated \rcFTL using four I/O traces generated from Sysbench~\cite{Sysbench} and Filebench~\cite{Filebench}.
\begin{table}[t]
	\centering
	\caption{I/O characteristics of traces used for evaluations.}
    \footnotesize
	\begin{tabular}{ c | c | c | c | c }
		\specialrule{.1em}{.05em}{.05em}
		            & {\sf OLTP} & {\sf NTRX} & {\sf Fileserver} & {\sf Varmail} \\
		\hhline{=====}
		 Read:Write & 7:3 & 0.5:9.5 & 4:6 & 4:6  \\
		\hline
		 WAF & 2.17 & 2.11 & 3.08 & 1.8 \\
		\specialrule{.1em}{.05em}{.05em}
	\end{tabular}
	\label{tab:workloads}%
\end{table}%
As shown in Table \ref{tab:workloads}, each workload has different
ratios between read and write and different WAF values.
Using these workloads, we evaluated the overall I/O throughput for three different \rcFTL versions, 
{\small{\sf rcFTL2}}, {\small{\sf rcFTL3}}, and {\small{\sf rcFTL4}}, 
where {\small{\sf rcFTL}}{\emph n} indicates that the maximum copyback threshold was set to {\emph n}.
All measurements were normalized over a page-level mapping FTL which always migrates data using the off-chip copy.

\subsection{Evaluation Results}

\begin{figure}[t]
	\centering
	\begin{subfigure}[t]{0.265\textwidth}
		\centering
		\includegraphics[width=1.0\linewidth]{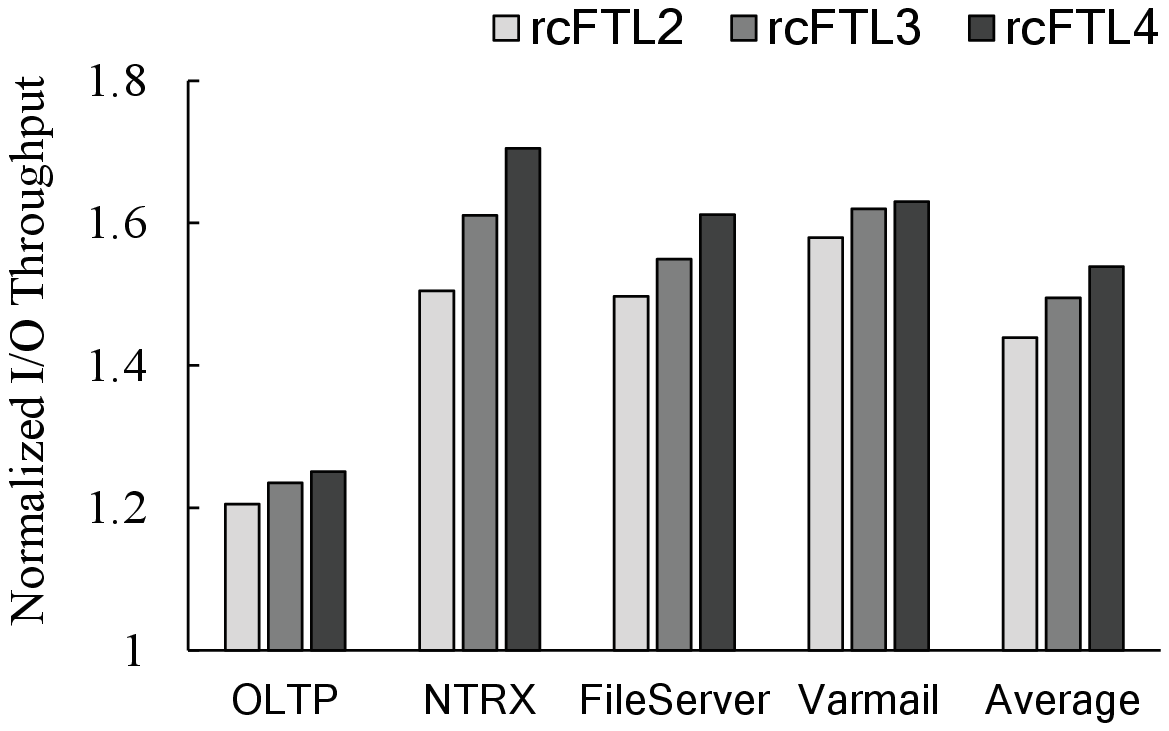}
		\caption{Normalized I/O throughput.}
	\end{subfigure}%
	\hfil
	\begin{subfigure}[t]{0.205\textwidth}
		\centering
		\includegraphics[width=1.0\linewidth]{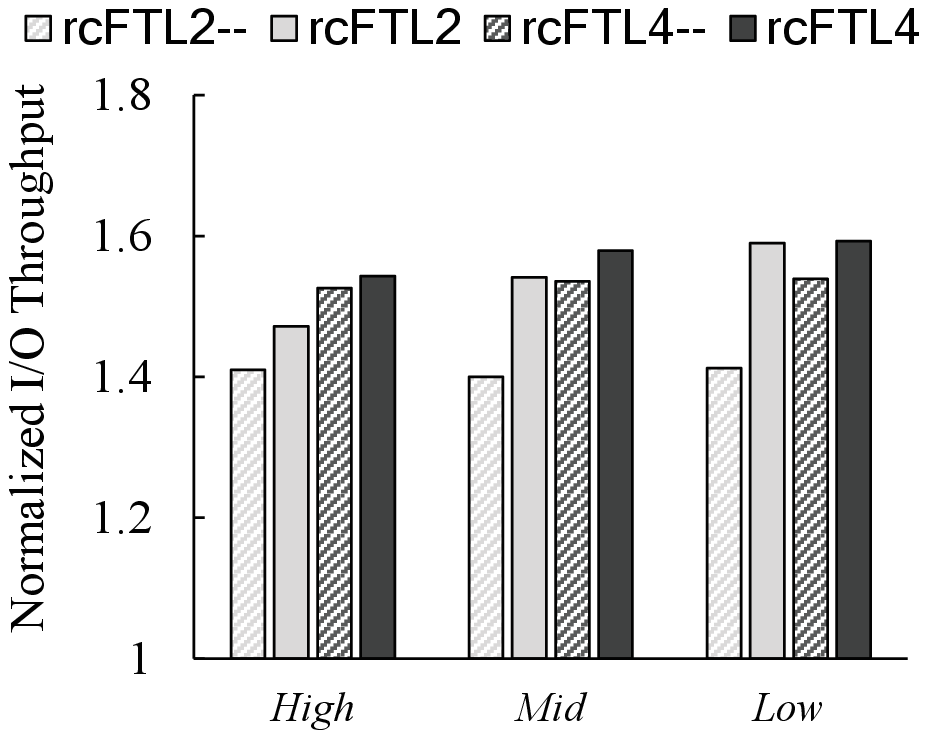}
		\caption{Effect of the mode selector.}
	\end{subfigure}%
	\caption{Performance comparisons of different \rcFTL versions.}
	\label{fig:performance}
\end{figure}
Fig. \ref{fig:performance}(a) shows normalized I/O throughputs of different \rcFTL versions.
As the copyback threshold increases, the I/O throughput increases accordingly 
because more data migrations are supported by \rcopyback.  
The overall I/O throughput was improved on average by 54\% 
in {\small{\sf rcFTL4}} over the baseline FTL.
Even  {\small{\sf rcFTL2}}, which can use \rcopyback only twice in a row, 
outperforms the baseline by 41\% on average.  
As the maximum copyback threshold increases, the I/O throughput of {\small{\sf NTRX}} quickly increases 
over other traces.
This difference comes from the difference in update patterns of each trace.  
In general, when data are updated sequentially (as in {\small{\sf Varmail}}), it is less likely that data are moved multiple times,
thus making {\small{\sf rcFTL}} with a higher maximum copyback threshold less efficient. 

In order to understand how the mode selector proposed in \rcFTL performs, 
we compared the performance of \rcFTL with {\small{\sf rcFTL}}{\emph {--}} (which uses \rcopyback in a greedy fashion).  
Fig.~\ref{fig:performance}(b) shows how these two \rcFTL versions compare under varying I/O intensity cases. 
In order to generate workload fluctuations, which are needed to properly evaluate the DMMS module, 
we generated three synthetic workloads, {\emph {High}}, {\emph {Mid}} and {\emph {Low}}, using Fio.
In {\emph {High}}, 70\% of I/O requests were issued without inter-request idle times while 30\% were issued with some idle times. 
For {\emph {Mid}} and {\emph {Low}}, the ratio between two requests is 50:50 and 30:70, respectively.
When the I/O intensity is lower, since the off-chip copy mode is more likely to be used in \rcFTL, 
\rcopyback-eligible blocks tend to increase over {\small{\sf rcFTL--}} 
because the per-block counters of more blocks are reset.
The increased number of \rcopyback-eligible blocks, in turn, improves the I/O throughput when the I/O intensity is high.
In Fig.~\ref{fig:performance}(b), {\small{\sf rcFTL2}} outperforms {\small{\sf rcFTL2--}} by this effect.  
In particular, for the {\emph {Low}} case, {\small{\sf rcFTL2}} improves the I/O throughput by 17\% over {\small{\sf rcFTL2--}}.

\section{Related Work}
There have been several studies to improve the performance 
of flash-based storage systems with the copyback operation. 
However, many existing techniques~\cite{seong2010hydra, abdurrab2013dloop, wang2015pcftl} 
are not applicable for modern NAND flash memory because they assumed 
an ideal SLC NAND flash memory where no error propagation occurs from successive copyback commands.
Other studies such as Jang {\it et al}.~\cite{chang2014efficient} considered 
the error propagation problem in their techniques.  
However, their solutions was to bring data out to the ECC module
to check the validity of data, thus minimizing the potential benefit of using copyback.   
Our technique is different from existing techniques in that 
the error propagation problem is fully controlled while maximizing the potential benefit of copyback.

\section{Conclusions}

We have presented \rcopyback to minimize performance degradations from internal data migrations in modern highly-parallel SSDs.   
From an experimental characterization study, we developed a \rcopyback operation model that takes as the key input the P/E cycle and data retention requirement.   
Based on the \rcopyback operation model, we have implemented a \rcopyback-aware FTL, \rcFTL, which intelligently manages when to use \rcopyback for a given I/O workload requirement.  Our experimental results show that \rcFTL can improve the overall I/O throughput by 54\% on average over an existing FTL with no copyback supported.


\bibliographystyle{ACM-Reference-Format}
\bibliography{copyback} 


\begin{thebibliography}{14}


\ifx \showCODEN    \undefined \def \showCODEN     #1{\unskip}     \fi
\ifx \showDOI      \undefined \def \showDOI       #1{#1}\fi
\ifx \showISBNx    \undefined \def \showISBNx     #1{\unskip}     \fi
\ifx \showISBNxiii \undefined \def \showISBNxiii  #1{\unskip}     \fi
\ifx \showISSN     \undefined \def \showISSN      #1{\unskip}     \fi
\ifx \showLCCN     \undefined \def \showLCCN      #1{\unskip}     \fi
\ifx \shownote     \undefined \def \shownote      #1{#1}          \fi
\ifx \showarticletitle \undefined \def \showarticletitle #1{#1}   \fi
\ifx \showURL      \undefined \def \showURL       {\relax}        \fi
\providecommand\bibfield[2]{#2}
\providecommand\bibinfo[2]{#2}
\providecommand\natexlab[1]{#1}
\providecommand\showeprint[2][]{arXiv:#2}

\bibitem[\protect\citeauthoryear{??}{Fil}{[n. d.]}]%
        {Filebench}
 \bibinfo{year}{[n. d.]}\natexlab{}.
\newblock \bibinfo{title}{Filebench}.
\newblock \bibinfo{howpublished}{{http://filebench.sourceforge.net}}.
  (\bibinfo{year}{[n. d.]}).
\newblock


\bibitem[\protect\citeauthoryear{??}{IDM}{[n. d.]}]%
        {IDM}
 \bibinfo{year}{[n. d.]}\natexlab{}.
\newblock \bibinfo{title}{{NAND Flash Performance Improvement Using Internal
  Data Move. Technical Note 29-15.}}
\newblock
  \bibinfo{howpublished}{{http://download.micron.com/pdf/technotes/nand/tn2915.pdf}}.
    (\bibinfo{year}{[n. d.]}).
\newblock


\bibitem[\protect\citeauthoryear{??}{Sys}{[n. d.]}]%
        {Sysbench}
 \bibinfo{year}{[n. d.]}\natexlab{}.
\newblock \bibinfo{title}{Sysbench}.
\newblock \bibinfo{howpublished}{{http://github.com/akopytov/sysbench}}.
  (\bibinfo{year}{[n. d.]}).
\newblock


\bibitem[\protect\citeauthoryear{Abdurrab et~al\mbox{.}}{Abdurrab
  et~al\mbox{.}}{2013}]%
        {abdurrab2013dloop}
\bibfield{author}{\bibinfo{person}{Abdul~R Abdurrab} {et~al\mbox{.}}}
  \bibinfo{year}{2013}\natexlab{}.
\newblock \showarticletitle{DLOOP: A flash translation layer exploiting
  plane-level parallelism}. In \bibinfo{booktitle}{{\em Proc. Int'l Symp.
  Parallel and Distributed Processing}}. \bibinfo{pages}{908--918}.
\newblock


\bibitem[\protect\citeauthoryear{Chang et~al\mbox{.}}{Chang
  et~al\mbox{.}}{2014}]%
        {chang2014efficient}
\bibfield{author}{\bibinfo{person}{Woo~Tae Chang} {et~al\mbox{.}}}
  \bibinfo{year}{2014}\natexlab{}.
\newblock \showarticletitle{An Efficient Copy-Back Operation Scheme Using
  Dedicated Flash Memory Controller in Solid-State Disks}.
\newblock \bibinfo{journal}{{\em Int'l Journal of Electrical Energy\/}}
  \bibinfo{volume}{2}, \bibinfo{number}{1} (\bibinfo{year}{2014}),
  \bibinfo{pages}{13--17}.
\newblock


\bibitem[\protect\citeauthoryear{Gupta et~al\mbox{.}}{Gupta
  et~al\mbox{.}}{2009}]%
        {gupta2009dftl}
\bibfield{author}{\bibinfo{person}{Aayush Gupta} {et~al\mbox{.}}}
  \bibinfo{year}{2009}\natexlab{}.
\newblock \showarticletitle{DFTL: a flash translation layer employing
  demand-based selective caching of page-level address mappings}. In
  \bibinfo{booktitle}{{\em Proc. Int'l Conf. Architectural Support for
  Programming Languages and Operating Systems}}. \bibinfo{pages}{229--240}.
\newblock


\bibitem[\protect\citeauthoryear{Jun et~al\mbox{.}}{Jun et~al\mbox{.}}{2015}]%
        {jun2015bluedbm}
\bibfield{author}{\bibinfo{person}{Sang-Woo Jun} {et~al\mbox{.}}}
  \bibinfo{year}{2015}\natexlab{}.
\newblock \showarticletitle{Bluedbm: An appliance for big data analytics}. In
  \bibinfo{booktitle}{{\em Proc. Int'l Symp. Computer Architecture}}.
  \bibinfo{pages}{1--13}.
\newblock


\bibitem[\protect\citeauthoryear{Kang et~al\mbox{.}}{Kang
  et~al\mbox{.}}{2006}]%
        {kang2006superblock}
\bibfield{author}{\bibinfo{person}{Jeong-Uk Kang} {et~al\mbox{.}}}
  \bibinfo{year}{2006}\natexlab{}.
\newblock \showarticletitle{A superblock-based flash translation layer for NAND
  flash memory}. In \bibinfo{booktitle}{{\em Proc. Int'l Conf. Embedded
  Software}}. \bibinfo{pages}{161--170}.
\newblock


\bibitem[\protect\citeauthoryear{Kim et~al\mbox{.}}{Kim et~al\mbox{.}}{2002}]%
        {kim2002space}
\bibfield{author}{\bibinfo{person}{Jesung Kim} {et~al\mbox{.}}}
  \bibinfo{year}{2002}\natexlab{}.
\newblock \showarticletitle{A space-efficient flash translation layer for
  CompactFlash systems}.
\newblock \bibinfo{journal}{{\em IEEE Trans. Consumer Electronics\/}}
  \bibinfo{volume}{48}, \bibinfo{number}{2} (\bibinfo{year}{2002}),
  \bibinfo{pages}{366--375}.
\newblock


\bibitem[\protect\citeauthoryear{Kim et~al\mbox{.}}{Kim et~al\mbox{.}}{2017}]%
        {kim2017improving}
\bibfield{author}{\bibinfo{person}{Myungsuk Kim} {et~al\mbox{.}}}
  \bibinfo{year}{2017}\natexlab{}.
\newblock \showarticletitle{Improving performance and lifetime of large-page
  NAND storages using erase-free subpage programming}. In
  \bibinfo{booktitle}{{\em Proc. Design Automation Conf.}}
\newblock


\bibitem[\protect\citeauthoryear{Lee et~al\mbox{.}}{Lee et~al\mbox{.}}{2016}]%
        {lee20167}
\bibfield{author}{\bibinfo{person}{Seungjae Lee} {et~al\mbox{.}}}
  \bibinfo{year}{2016}\natexlab{}.
\newblock \showarticletitle{A 128Gb 2b/cell NAND flash memory in 14nm
  technology with tPROG= 640$\mu$s and 800MB/s I/O rate}. In
  \bibinfo{booktitle}{{\em Proc. Int'l Solid-State Circuits Conf.}}
  \bibinfo{pages}{138--139}.
\newblock


\bibitem[\protect\citeauthoryear{Lee et~al\mbox{.}}{Lee et~al\mbox{.}}{2006}]%
        {lee2006fast}
\bibfield{author}{\bibinfo{person}{Sang-Won Lee} {et~al\mbox{.}}}
  \bibinfo{year}{2006}\natexlab{}.
\newblock \showarticletitle{FAST: An efficient flash translation layer for
  flash memory}. In \bibinfo{booktitle}{{\em Proc. Int'l Conf. Embedded and
  Ubiquitous Computing}}. \bibinfo{pages}{879--887}.
\newblock


\bibitem[\protect\citeauthoryear{Seong et~al\mbox{.}}{Seong
  et~al\mbox{.}}{2010}]%
        {seong2010hydra}
\bibfield{author}{\bibinfo{person}{Yoon~Jae Seong} {et~al\mbox{.}}}
  \bibinfo{year}{2010}\natexlab{}.
\newblock \showarticletitle{Hydra: A block-mapped parallel flash memory
  solid-state disk architecture}.
\newblock \bibinfo{journal}{{\em IEEE Trans. Computers\/}}
  \bibinfo{volume}{59}, \bibinfo{number}{7} (\bibinfo{year}{2010}),
  \bibinfo{pages}{905--921}.
\newblock


\bibitem[\protect\citeauthoryear{Wang and Xie}{Wang and Xie}{2015}]%
        {wang2015pcftl}
\bibfield{author}{\bibinfo{person}{Wei Wang} {and} \bibinfo{person}{Tao Xie}.}
  \bibinfo{year}{2015}\natexlab{}.
\newblock \showarticletitle{PCFTL: A plane-centric flash translation layer
  utilizing copy-back operations}.
\newblock \bibinfo{journal}{{\em IEEE Trans. Parallel and Distributed
  Systems\/}} \bibinfo{volume}{26}, \bibinfo{number}{12}
  (\bibinfo{year}{2015}), \bibinfo{pages}{3420--3432}.
\newblock


\end{thebibliography}

\end{document}